\documentclass[10pt, conference, compsocconf]{IEEEtran}
\pdfoutput=1

\usepackage{bm,mathtools} 
\usepackage[binary-units=true]{siunitx} 
\usepackage{array,makecell,multirow,booktabs} 
\usepackage{todonotes} 
\usepackage{algorithmicx,algorithm} 
\usepackage[noend]{algpseudocode}
\usepackage{listings} 
\usepackage{mdframed} 
\usepackage{graphicx}
\usepackage[subrefformat=parens]{subcaption} 
\usepackage{adjustbox}
\usepackage{xcolor, soul}

\usepackage{fancyhdr}
\usepackage[normalem]{ulem}
\usepackage[hyphens]{url}
\usepackage[sort,nocompress]{cite}
\usepackage[final]{microtype}
\usepackage{flushend}

\usepackage[bookmarks=true,breaklinks=true,letterpaper=true,colorlinks,linkcolor=black,citecolor=blue,urlcolor=black]{hyperref}
\hypersetup{draft} 

\newcommand\name{GRIP}

\DeclareSIUnit\OP{OP}
\DeclareSIUnit\FLOP{Flop}

\algnewcommand\algorithmicinput{\textbf{Input:}}
\algnewcommand\Input{\item[\algorithmicinput]}
\algnewcommand\algorithmicoutput{\textbf{Output:}}
\algnewcommand\Output{\item[\algorithmicoutput]}

\mdfdefinestyle{mdtight}{
    skipabove=0pt,
    skipbelow=0pt,
    leftmargin=0pt,
    rightmargin=0pt,
    innerleftmargin=2pt,
    innerrightmargin=2pt,
    innertopmargin=0pt,
    innerbottommargin=0pt
}

\definecolor{black}{RGB}{0,0,0}
\definecolor{green}{RGB}{26,128,26}

\lstdefinestyle{greta}{
    basicstyle=\footnotesize,
    keywordstyle=\color{blue}\ttfamily,
    stringstyle=\color{red}\ttfamily,
    commentstyle=\small\color{green}\ttfamily,
    morekeywords={class,def,return},
    morecomment=[l]{//},
    breaklines=true
}

\title{\name{}: A Graph Neural Network Accelerator Architecture} 

\author{
    \IEEEauthorblockN{
        Kevin Kiningham\IEEEauthorrefmark{1}, Christopher Ré, Philip Levis
    }
    \IEEEauthorblockA{Stanford University}
    \IEEEauthorrefmark{1}\normalfont{\texttt{kkiningh@stanford.edu}}
}

\begin{document}
\maketitle
\pagestyle{plain}


\begin{abstract}

We present \name{}, a graph neural network accelerator architecture designed for low-latency inference.
Accelerating GNNs is challenging because they combine two distinct types of computation: arithmetic-intensive \textit{vertex-centric} operations and memory-intensive \textit{edge-centric} operations.
\name{} splits GNN inference into a fixed set of edge- and vertex-centric execution phases that can be implemented in hardware.
We then specialize each unit for the unique computational structure found in each phase.
For vertex-centric phases, \name{} uses a high performance matrix multiply engine coupled with a dedicated memory subsystem for weights to improve reuse.
For edge-centric phases, \name{} use multiple parallel prefetch and reduction engines to alleviate the irregularity in memory accesses.
Finally, \name{} supports several GNN optimizations, including a novel optimization called vertex-tiling which increases the reuse of weight data.

We evaluate \name{} by performing synthesis and place and route for a \SI{28}{\nano\meter} implementation capable of executing inference for several widely-used GNN models (GCN, GraphSAGE, G-GCN, and GIN).
Across several benchmark graphs, it reduces 99th percentile latency by a geometric mean of {$17\times$} and {$23\times$} compared to a CPU and GPU baseline, respectively, while drawing only \SI{5}{\watt}.
\end{abstract}

\begin{IEEEkeywords}
Deep Learning; Hardware Acceleration; Algorithm-Hardware co-Design; ASIC;
\end{IEEEkeywords}

\section{Introduction}
\label{sec:introduction}

Traditional deep neural networks (DNNs) rely on regularly structured inputs (e.g. vectors, images, or sequences)  making them difficult to use in domains where data is naturally irregular (e.g. user connections on social media).
Graph neural networks (GNNs) tackle this limitation by extending DNNs to allow arbitrarily structured graph-valued inputs, where feature vectors are associated with the edges and vertices of a graph\footnote{
Following the convention in prior work~\cite{hamilton2017inductive}, for clarity we call a GNN's input a \emph{graph} and the GNN itself a \emph{network}.
}.
GNNs have found significant success in a range of practical tasks, including surfacing related content on social media~\cite{ying2018graph}, recommending meals on delivery platforms~\cite{uber2020gnn}, and improving circuit testability for EDA~\cite{ma2019high}.

GNNs combine two distinct types of operations~\cite{gilmer2017neural,ma2019neugraph}: \emph{vertex-centric}, which are associated with graph vertices, and \emph{edge-centric}, which are associated with edges.
Vertex-centric operations are computationally regular and primarily consist of multiplying vertex feature vectors by large weight matrices.
These weights are shared across all vertices, leading to significant opportunities for data reuse. 
Edge-centric operations are similar to those found in graph analytics (e.g. neighborhood reduction~\cite{wang2017gunrock}).
Their computational structure depends on the often sparse and irregular structure of the input graph.
This results in many random memory accesses and limited data reuse, but also requires relatively little computation.

The combination of these two types of computation makes GNN inference inefficient on existing architectures.
As a result, GNNs have much higher inference latency than other neural networks, limiting them to applications where inference can be pre-computed offline~\cite{ying2018graph}.
Most DNN accelerators (e.g. the TPU~\cite{jouppi2017datacenter}) are optimized for dense, regular computation, making edge operations hard to implement efficiently~\cite{balog2019fast}.
Graph analytics accelerators (e.g Graphicionado~\cite{ham2016graphicionado}) are designed for workloads that require little computation per-vertex and have difficulty exploiting data reuse in vertex-centric operations.
Prior work has demonstrated inference on CPUs and GPUs is limited by architectural issues, such as cache and memory bandwidth bottlenecks~\cite{yan2020hygcn, gunrock2019graphsage}.

This paper proposes \name{} (GRaph Inference Processor), an accelerator architecture designed for low-latency GNN inference.
\name{}'s programming model is inspired by GReTA~\cite{recoml20greta}, a decomposition of GNN inference into a fixed set of edge- and vertex-centric phases.
\name{} implements each phase with separate specialized on-chip memory and execution units.
For example, \name{} alleviates irregularity in the \textit{edge-accumulate} phase by using multiple parallel prefetch engines to load data.
This allows \name{} to support a broader class of GNNs than prior work, including emerging models that perform complex computation per-edge.
Finally, \name{} includes hardware support for several optimizations: caching partitions of feature data, inter-phase pipelining, and preloading weights between layers.
We also introduce a novel GNN optimization called vertex-tiling that substantially improves latency by increasing the reuse of weight values during inference.

\subsection{Contributions}
This paper makes the following contributions:

\begin{enumerate}
    \item \name{}, an accelerator architecture for low-latency GNN inference. \name{} is efficient across a wide range of models and has numerous hardware optimizations to improve inference latency.
    \item A novel optimization for GNN inference called vertex-tiling, which improves performance by increasing reuse of weights.
    \item A detailed description of a \SI{28}{\nano\meter} implementation of \name{} capable of executing four representative GNNs (GCN, GraphSAGE, G-GCN, and GIN).
    Evaluated across several benchmark graphs, our implementation reduces 99th percentile latency by a geometric mean of {$17\times$} and {$23\times$} compared to an Intel Xeon CPU and Nvidia P100 GPU baseline, respectively.
\end{enumerate}

\section{Background and Motivation}
\label{sec:background}

\begin{figure}[tb]
    \centering
    \subcaptionbox{Input graph.\label{subfig:example-graph}}[.4\linewidth]{
        \includegraphics{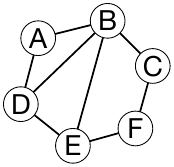}
    }
    \hfill
    \subcaptionbox{Nodeflow.\label{subfig:example-nodeflow}}[.59\linewidth]{
        \includegraphics{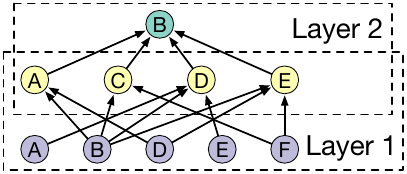}
    }
    \vskip \baselineskip
    \subcaptionbox{Inference dataflow.\label{subfig:example-inference}}{
        \includegraphics{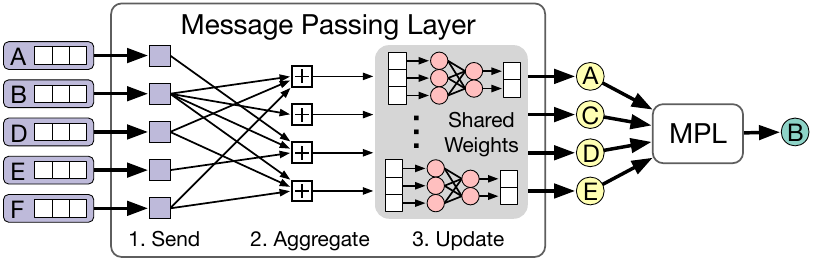}
    }
    \vskip \baselineskip
    \caption{An example of performing GCN inference on vertex B with two layers. The nodeflow~\subref{subfig:example-nodeflow} describes the propagation of features during inference~\subref{subfig:example-inference}.}
    \label{fig:intro-example}
\end{figure}

\subsection{Graph Neural Networks}
\label{sec:gnn-background}

GNNs~\cite{battaglia2018relational, wu2019comprehensive} are a class of DNN that operate on graph-valued data.
Unlike traditional DNNs, GNNs directly take advantage of graph structure during learning and inference.
For example, consider the task of classifying web-pages by topic.
A pure content approach (e.g. a classic recurrent neural network) considers only features derived from a page's content.
However, a significant amount of information is stored in the structure of links \emph{between} pages.
By modeling these links as a graph, a GNN can natively leverage both page content and link structure.
GNN-based methods have achieved state of the art performance on a diverse set of graph-related tasks, including link prediction~\cite{zhang2018link}, vertex classification~\cite{ying2018graph}, and clustering~\cite{ying2018hierarchical}.

\textbf{Message-Passing Layer}.
Modern GNNs are typically composed of multiple message-passing layers~\cite{gilmer2017neural}, shown in Alg.~\ref{alg:gnn-forward}.
The layer takes as input a graph $G$ consisting of a set of vertices $V$ and edges $E$.
Each vertex and edge is assigned a feature vector $\bm{h}_v$ and $\bm{h}_{(u,v)}$ respectively.
Computation is split into three operations:
\begin{itemize}
    \item \textit{Send} computes a message vector $\bm{m}_{u,v}$ for each edge.
    \item \textit{Aggregate} reduces incoming messages for each vertex to a vector $\bm{a}_v$. The neighborhood function $N(v)$ determines which messages are considered, typically using a fixed size random sample.
    \item \textit{Update} combines each vertex's current value with the output of aggregation to produce an updated vector $\bm{z}_v$.
\end{itemize}
By iteratively applying $K$ of these layers, the final state for each vertex captures information about the structure of its $K$-hop neighborhood.

\begin{algorithm}[tb]
\caption{Message Passing Layer Forward Pass}
\label{alg:gnn-forward}
\begin{algorithmic}[1]
\Input Graph $G = (V,\,E)$; Vertex and edge features $\bm{h}_v$, $\bm{h}_{(u,v)}$
\Output Updated vertex features $\bm{z}_v$
\For{$(u,\,v)$ \textbf{in} $E$}
    \State $\bm{m}_{u,v} \leftarrow \mathrm{Send}(\bm{h}_{v},\,\bm{h}_{u},\,\bm{h}_{(u,v)})$
\EndFor
\For{$v$ \textbf{in} $V$}
    \State $\bm{a}_v \leftarrow \mathrm{Aggregate}(\{ \bm{m}_{u,v} \mid u \in N(v) \})$
    \State $\bm{z}_v \leftarrow \mathrm{Update}(\bm{h}_v,\,\bm{a}_v)$
\EndFor
\end{algorithmic}
\end{algorithm}

\textbf{Pooling and Readout}.
Two other layer types are also used in some GNNs.
\textit{Pooling} defines a method of coarsening a graph by combining the features of clusters of vertices.
\textit{Readout} is a special case of pooling which uses a single graph-wide cluster to produce a representation for an entire graph (e.g. for graph classification).
In this paper, we treat both layers as slightly modified versions of message-passing, where edges connect vertices to clusters rather than other vertices.

\textbf{Nodeflow}.
A nodeflow~\cite{huang2019nodeflow} is a bipartite data structure that describes how features are propagated during message-passing.
It is typically generated during a preprocessing step before inference, but can also be created on-demand (e.g. for dynamic graphs).
The nodeflow is most useful when performing inference on a subset of the graph since it makes it easy to determine which edges and vertices are required to update a specific vertex.
It can also be used to separate sampling from inference by precomputing the neighborhood function and encoding the result directly in the nodeflow.

In this paper, we denote the nodeflow for a particular layer as the three-tuple $(U, V, E)$, where $U$ is the set of vertices read during inference, $V$ is the set of vertices updated, and $E$ is a set of edges connecting vertices in $U$ to $V$.
Fig.~\ref{fig:intro-example} shows an example of using the nodeflow to compute inference in a two layer GNN.

\textbf{GCN}.
We use the Graph Convolutional Network~\cite{kipf2017semi} (GCN) as a concrete running example of a GNN model.
GCN uses multiple message-passing layers with the following send, aggregate, and update operations
\begin{align*}
    \bm{m}_{u,v} &\leftarrow \bm{h}_u \\
    \bm{a}_v &\leftarrow \mathrm{mean}(\{\bm{m}_{u,v} \mid u \in N(v)\}) \\
    \bm{z}_v &\leftarrow \mathrm{ReLU}(W \bm{a}_v )
\end{align*}
where $W$ is a trainable weight matrix.
We can rewrite this to use sparse-dense matrix multiplication (SpMM)
\begin{equation}
    Z \leftarrow \mathrm{ReLU}(\hat{A} H W)
    \label{eq:gcn-optimized}
\end{equation}
where $\hat{A}$ is a sparse matrix derived from the nodeflow and $H$ and $Z$ are dense matrices formed from the set of input and output features respectively.
This allows GCN inference to be implemented using operations from highly optimized sparse matrix libraries, such as Intel MKL~\cite{intel2019mkl} or cuSPARSE~\cite{nvidia2019cusparse}.

\subsection{Performance Challenges of GNNs}
\label{sec:performance-challenges}

\begin{figure}[thb]
    \centering
    \includegraphics{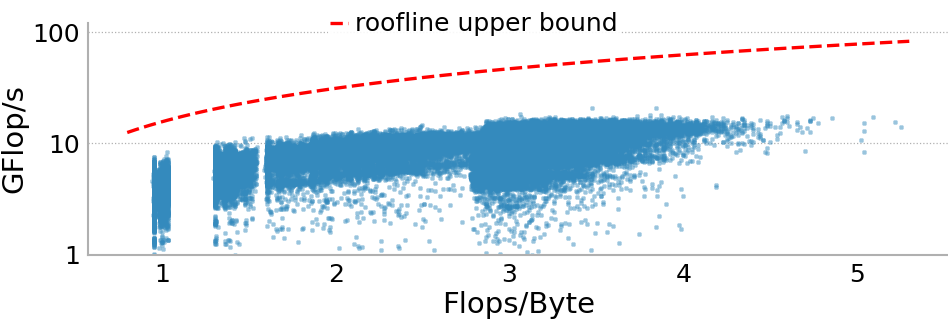}
    \vskip 0.5em
    \caption{CPU performance of GCN inference for vertices in the Pokec~\cite{snapnets} dataset. Bottlenecks in cache bandwidth result in a significant gap between measured performance and the roofline upper bound.}
    \label{fig:gnn-roofline}
\end{figure}

To demonstrate the performance challenges of GNNs in practice, we implement 2-layer GCN using the SpMM form in Eq.~\ref{eq:gcn-optimized}.
Our implementation uses Tensorflow compiled with Intel MLK run on a single socket of an Intel Xeon E5-2690v4.
In Fig.~\ref{fig:gnn-roofline}, we plot measured performance verses arithmetic intensity for each vertex in the Pokec dataset.
Arithmetic intensity depends on the number of unique neighbors that must be read during inference, which is determined by the local graph structure of each vertex.

In this dataset, inference performance is theoretically bottlenecked by off-chip memory bandwidth for all vertices.
However, there is a significant gap between the theoretical upper bound and the actual measured performance at higher levels of arithmetic intensity.
Profiling shows the primary bottleneck is last level cache bandwidth, a result consistent with prior analysis of GPU performance~\cite{gunrock2019graphsage}.
In our experiment, the highest arithmetic intensities occur when a vertex appears in multiple neighborhoods and its feature vector can be reused.
However, if multiple cores are reading or writing a vertex in parallel, this also results in higher utilization of cache bandwidth.
Additionally, features must compete with large weight values that also occupy the cache and consume bandwidth during the vertex-centric \textit{Update} operation.

\textbf{Opportunities for Acceleration}.
While GNN inference performance may be limited on existing hardware, the difficulties described in this section can be overcome with a custom architecture.
In particular, we propose using separate specialized memory and execution units for each edge- and vertex-centric operation.
To specialize for vertex-centric operations, we use a dedicated high performance matrix-multiplication unit.
Weights are stored on-chip in dedicated memory with a level of caching to improve reuse.
For edge-centric operations, we prefetch data for multiple edges in parallel and specialize the on-chip feature memory to enable fast gather and reduction operations. 
Finally, since the nodeflow is known statically, we can also improve off-chip access efficiency by scheduling bulk transfers of feature data rather than loading on demand during execution.
Taken together, this gives a significant opportunity for improving GNN inference performance.

\section{Related Work}

\textbf{DNN Accelerators}.
A significant number of custom neural network accelerates have been developed, mostly focused on dense operations~\cite{chen2014dadiannao, chen2014diannao, zhang2015optimizing, chen2017eyeriss, farabet2011neuflow, jouppi2017datacenter, lu2017flexflow, du2015shidiannao, gao2017tetris, venkataramani2017scaledeep, gao2019tangram}.
However, edge-centric operations are difficult to implement efficiently on these architectures~\cite{balog2019fast}, which are much more computationally irregular than traditional DNNs.
\name{} natively supports edge-centric operations by using a graph-processing based programming model (Sec.~\ref{sec:greta}) and by a combination of specialized memory for edge accesses and software techniques.
In Sec.~\ref{sec:eval-prior-work} we estimate \name{} to be $2.4\times$ faster than a comparable TPU-like accelerator modified specifically to improve GNN inference performance.

\textbf{GCN Accelerators}.
HyGCN~\cite{yan2020hygcn} and GraphACT~\cite{zeng2020graphact} are two accelerators designed to for graph convolutional networks, a subclass of graph neural networks.
Like \name{}, these accelerators use separate edge- and vertex-centric units for GNN computation.
\name{} builds on these designs by handling a much more general set of GNNs that includes models that use computation associated with edges.
This is important for many emerging state-of-the-art GNNs, such as Graph Attention Networks~\cite{velivckovic2017graph}.
Additionally, \name{}'s support for vertex-tiling reduces the amount of weight bandwidth required during vertex-oriented operations.
In Sec.~\ref{sec:eval-prior-work} we estimate this improves performance by $4.5\times$ compared to HyGCN.

\textbf{Graph Analytics Accelerators}.
Specialized accelerators have also been proposed for graph analytics workloads~\cite{ozdal2017graph, nurvitadhi2014graphgen, oguntebi2016graphops}.
However, these workloads require relatively little computation per-vertex and typically use scalars rather than large feature vectors.
Thus, the computation and memory access patterns are very different.
In Sec.~\ref{sec:eval-prior-work}, we estimate \name{} to be $8.1\times$ faster than the approach of Graphicionado~\cite{ham2016graphicionado}.

\textbf{GNN Optimizations}.
Many optimizations have been proposed to improve GNN performance.
Common techniques include scheduling computation to reducing the impact of sparsity~\cite{ma2019neugraph, balog2019fast, chiang2019cluster}, improved sampling~\cite{chen2018fastgcn}, or eliminating redundant computation~\cite{zeng2020graphact}.
These techniques are compatible with \name{} and can be used for additional performance.

\section{Programming Model}
\label{sec:greta}

\name{}'s programming model is based on GReTA~\cite{recoml20greta}, a graph-processing abstraction specialized for implementing GNNs.
GReTA decomposes GNN layers into four stateless user-defined functions (UDFs): \texttt{\underline{g}ather}, \texttt{\underline{re}duce}, \texttt{\underline{t}ransform}, and \texttt{\underline{a}ctivate}.
\name{} invokes each UDF in one of three execution phases: \textit{edge-accumulate}, \textit{vertex-accumulate}, and \textit{vertex-update}.
\name{} also allows programs to be composed by using the result of one program as the features or accumulator in another.
This flexibility allows a wide range of GNN models to be implemented.

\begin{algorithm}[t]
\caption{\name{} Program Execution Semantics}
\label{alg:greta-semantics}
\begin{algorithmic}[1]
\Input Layer nodeflow $(U, V, E)$; Vertex data $\bm{h}_u$ and $\bm{h}_v$; Edge data $\bm{h}_{(u, v)}$; Accumulators $\bm{e}_v$ and $\bm{a}_{v}$; Weights and biases $W$
\Output Updated vertex data $\bm{z}_v$
\State \texttt{/* Edge-Accumulate Phase */}
\For{$(u, v)$ \textbf{in} $E$}
    \State $\bm{e}_v$ = reduce($\bm{e}_v$, gather(\upshape $\bm{h}_u$, $\bm{h}_v$, $\bm{h}_{(u, v)}$))
\EndFor
\State \texttt{/* Vertex-Accumulate Phase */}
\For{$v$ \textbf{in} $V$}
    \State $\bm{a}_v$ = transform($\bm{a}_v$, $\bm{e}_v$, $W$)
\EndFor
\State \texttt{/* Vertex-Update Phase */}
\For{$v$ \textbf{in} $V$}
    \State $\bm{z}_v$ = activate($\bm{a}_v$)
\EndFor
\end{algorithmic}
\end{algorithm}

\textbf{Data Model}.
UDFs are restricted in the types of data they can access to in order to simplify hardware implementation.
\name{} programs use four types of data:
(1) A nodeflow $NF = (U, V, E)$ encodes computational structure by defining the vertices and edges to read and update.
(2) Feature vectors $\bm{h}_u$, $\bm{h}_v$, and $\bm{h}_{(u,v)}$ associated with nodeflow input vertices, output vertices, and edges respectively.
(3) A set of constant layer weights $W$.
(4) Edge-accumulator $\bm{e}_v$ and vertex-accumulator $\bm{a}_v$ associated with each output vertex.

\textbf{Execution Semantics}.
UDFs are executed in three phases:
\begin{enumerate}
    \item \textit{Edge-accumulate} iterates over nodeflow edges and invokes \texttt{gather} and \texttt{reduce}.
    \texttt{Gather} reads features associated with an edge to produce a message value.
    \texttt{Reduce} accumulates messages sharing an output vertex into $\bm{e}_v$.
    This results in a single value per output vertex.
    \item \textit{Vertex-accumulate} iterates over each output vertex and combines $\bm{e}_v$ with the previous accumulator state $\bm{a}_v$ using \texttt{transform}.
    \texttt{Transform} is the only UDF with access to layer weights and is usually the most computationally expensive operation in a layer (e.g. matrix multiplication).
    \item \textit{Vertex-update} again iterates over each output vertex and applies \texttt{activate} to $\bm{a}_v$.
    The \texttt{activate} UDF typically implements the non-linear operations required in a layer (e.g. the activation function).
    This produces a final updated value for each vertex $\bm{z}_v^\prime$.
\end{enumerate}
Alg.~\ref{alg:greta-semantics} shows the full execution semantics of a \name{} program.

\subsection{Layer Implementation}

\begin{figure}[t]
    \centering
    {\includegraphics{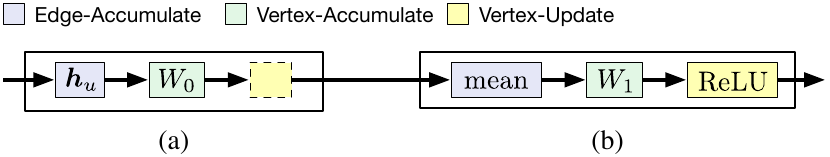}\phantomsubcaption\label{subfig:greta-split1}\phantomsubcaption\label{subfig:greta-split2}}
    \vskip \baselineskip
    \caption{
        Modifying the GCN \textit{Send} operation (Eq.~\ref{eq:gcn-modified}) requires splitting the layer into two sequential programs \subref{subfig:greta-split1} and \subref{subfig:greta-split2}.
        The dashed box in \subref{subfig:greta-split1} indicates a phase with no computation.
    }
    \label{fig:greta-split}
\end{figure}

\begin{figure}[t]
    \centering
    \subcaptionbox{GraphSAGE~\cite{hamilton2017inductive}\label{subfig:greta-examples-gsmax}}{
      \includegraphics{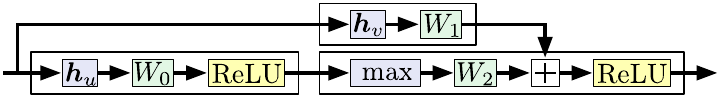}
    }
    \vskip 0.5\baselineskip
    \subcaptionbox{GIN~\cite{xu2018powerful}\label{subfig:greta-examples-gin}}{
      \includegraphics{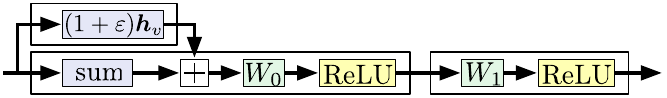}
    }
    \vskip 0.5\baselineskip
    \subcaptionbox{G-GCN~\cite{bresson2017residual, marcheggiani2017encoding, allamanis2018learning}\label{subfig:greta-examples-ggcn}}{
      \includegraphics{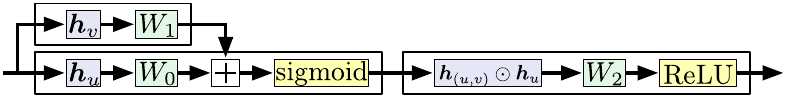}
    }
    \vskip \baselineskip
    \caption{\name{} implementation of several GNN models. Plus-boxes indicate the output of one program is used as the edge or vertex-accumulator of another. Phases with no associated computation are omitted.}
    \label{fig:greta-example}
\end{figure}

The decomposition used by \name{} is expressive enough to allow implementing a wide variety of GNNs.
Implementing a particular layer is typically straightforward since each phase naturally maps to the operations of the massage-passing layer introduced in Sec.~\ref{sec:background}.
However, some complex models may require a layer to be split into multiple programs.
This is especially true for models that require significant computation per-edge.
For example, consider the following modified GCN \textit{Send} operation
\begin{equation}
    \bm{m}_{u,v} \leftarrow W_0 \bm{h}_u \label{eq:gcn-modified}
\end{equation}
This cannot be mapped directly to \texttt{gather} and \texttt{reduce} since they do not have access to the layer weights.
Instead, we implement this layer by splitting it into two \name{} programs as shown in Fig.~\ref{fig:greta-split}.
Note that splitting a layer may result in each program iterating over a different nodeflow.
For example, the program in Fig.~\ref{subfig:greta-split1} iterates over an identity nodeflow where all vertices are only self-connected.
In Fig.~\ref{fig:greta-example}, we demonstrate the flexibility of this approach by showing the implementation of a variety of different GNN models.

\section{The \name{} Architecture}
\label{sec:grip-architecture}

\begin{figure*}[thb]
    \centering
    \includegraphics{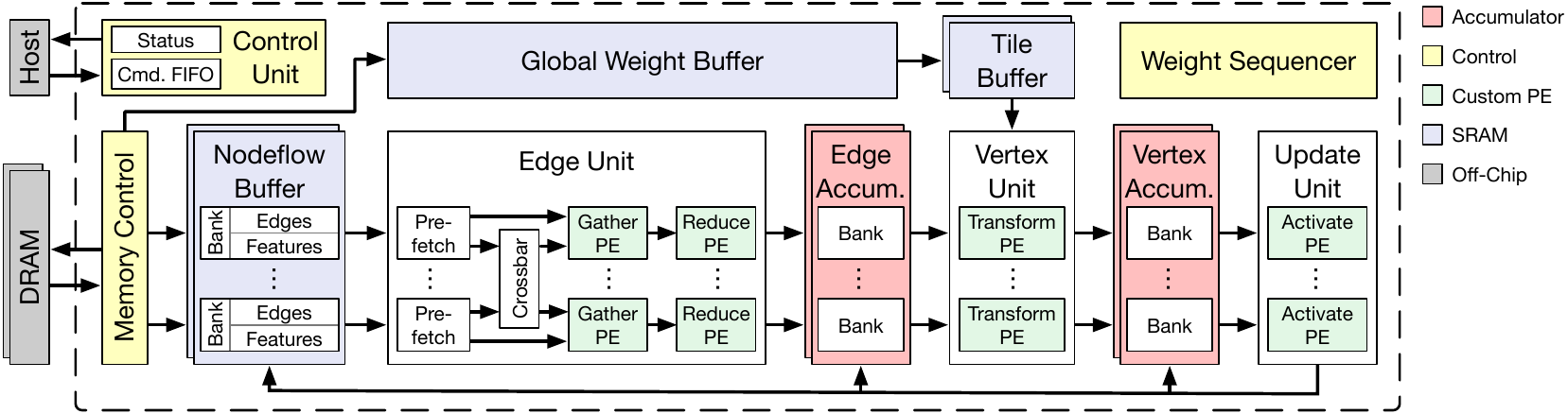}
    \vskip \baselineskip
    \caption{High-level overview of \name{}.}
    \label{fig:architecture}
\end{figure*}

\name{} is an accelerator architecture for low-latency GNN inference.
Rather than designing around a specific GNN, \name{} allows users to customize the architecture by implementing four processing elements (PEs) corresponding to each UDF of GReTA.
This allows \name{} to be used to accelerate a wide variety of models.
In this section, we describe an overview of \name{} and the microarchitecture of each execution unit.
A high level overview of \name{} is shown in Fig.~\ref{fig:architecture}.

\subsection{Overview}

\textbf{Control}.
\name{} is controlled by a host system that sends commands to execute different operations or transfer data.
The control unit dequeues each command in-order and issues them asynchronously to individual execution units or the memory controller.
Additionally, almost all buffers use double-buffering to allow overlapping the execution of different operations with moving data between buffers or loading from off-chip.
A barrier command is used to enforce dependencies by preventing new commands from being issued until all previous commands have completed.
Each command also updates a global status register on completion, which can be queried by the host to monitor execution.


\textbf{Execution Units}.
\name{} has three core execution units: the edge unit, the vertex unit, and the update unit.
The edge unit performs the edge-accumulate phase by iterating over the edges of the nodeflow, which is stored in the nodeflow buffer.
The edge unit then reads the associated features, executes \texttt{gather}, and finally accumulates the result into the edge accumulator using \texttt{reduce}.

The vertex unit performs the vertex-accumulate phase by iterating over the output vertices corresponding to the accumulated edge values.
It then executes transform, accumulating the result into the vertex accumulator.
The vertex unit also reads weight values from the tile buffer, which caches tiles of weight values from the global weight buffer.
To synchronize the tile buffer and the vertex unit, the weight sequencer controls iterating over the tiles as described in Section~\ref{sec:tiling}.
Since weight values are shared across all nodeflow output vertices, the global weight buffer is only required to be loaded once at the beginning of a \name{} program.

Finally, the update unit performs the vertex-update phase by reading the accumulated values for each vertex and passing the values to the activate PE.
The result is written to the nodeflow buffer as an updated feature, or to the edge or vertex accumulator.
This allows efficiently passing values between different \name{} programs when they are executed in sequence.

\textbf{PE Implementation}.
\name{} allows users to customize four PEs corresponding to the UDFs introduced in Sec.~\ref{sec:greta}.
These can be implemented in multiple ways depending on the user's needs.
For example, they could be implemented using a reconfigurable fabric (e.g. an FPGA) for maximum flexibility.
Alternatively, they could be implemented using a model specific circuit to optimize for area or performance.

Our implementation uses a programmable ALU based approach.
Since most common GNNs only require a small number of operations in practice, this allows us to support a range of models on the same hardware while remaining reasonably efficient in practice.
Specifically, we allow \texttt{gather} to be identity (e.g. $\bm{h}_u$ or $\bm{h}_v$), element-wise sum, product, or scale by constant; \texttt{reduce} to be element-wise sum, max, or mean; \texttt{transform} to be matrix multiplication followed by element-wise sum; and \texttt{activate} to be either ReLU or a LUT operation which we describe in Sec.~\ref{sec:activation}.
While these cover most GNN models we investigated, expanding the set of supported operations may be required for other GNNs.
We leave exploring other possible implementations for future work.

\textbf{Memory Controller}.
The memory controller is responsible for moving data on- and off-chip.
Instead of each execution unit issuing requests to the memory controller directly, the host is required to statically schedule memory transfers before execution.
This is possible since the set of features required for inference can be easily determined from the nodeflow.
This also prevents individual units from stalling on external memory accesses, but requires scheduling commands such that loading data fully overlaps with execution (Sec.~\ref{sec:partitioning}).

\subsection{Edge Unit}

\begin{figure}[thb]
    \centering
    \includegraphics{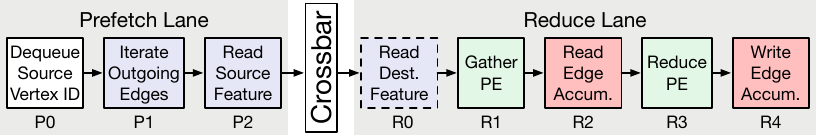}
    \vskip \baselineskip
    \caption{The edge unit pipeline is split into source-oriented (P0-P2) and destination-oriented (R0-R4) sections called lanes that can be independently replicated. Stage R0 is only used for models that require reading source features.}
    \label{fig:edge-unit}
\end{figure}

The edge unit pipeline is split into two distinct halves (Fig.~\ref{fig:edge-unit}).
Stages P0-P2 implement \textit{prefetch}, which iterates over the edges of the nodeflow and reads the features corresponding to the source vertex.
The result is passed to \textit{reduce} (stages R0-R4), which reads the corresponding destination feature and then applies \texttt{gather} and \texttt{reduce}, accumulating the result into the edge accumulator.
\name{} allows optionally disabling stage R0 since most models do not require reading source features.

\textbf{Parallelization}.
The edge-accumulator value for each output vertex can be computed independently.
This means there is a significant amount of parallelism that can be exploited to improve the performance of the edge unit.
A simple method to parallelize execution is to duplicate the elements of the edge unit into $N$ identical copies.
Each copy can then be assigned a subset of output vertices to process in parallel (e.g. by a random hash of the vertex ID).
However, since the nodeflow buffer is read every cycle by each lane, this approach requires adding $2N$ ports to the nodeflow buffer.

Instead, \name{} duplicates prefetch and reduce into $N$ and $M$ copies called lanes.
Each lane is statically assigned a partition of input vertices (for prefetch) or output vertices (for reduce).
Similarly, edges are assigned to a prefetch lane based on the edge's source vertex.
During execution, each prefetch lane iterates over its assigned edges and reads each edge's corresponding feature.
It then sends the feature data through an $N \times M$ crossbar to the reduce lane assigned to the destination vertex.
This design restricts each lane to accessing only its assigned subset of features and edges, allowing \name{} to partition the nodeflow buffer into $N + M$ separate SRAMs.
As a result, \name{} scales to a much larger number of lanes than the simpler design.
Additionally, our implementation of \name{} extends this scheme to include off-chip memory by storing feature data pre-partitioned and setting the number of prefetch lanes equal to the number of DRAM channels.

\subsection{Vertex Unit}


The vertex unit implements the vertex-accumulate phase by iterating over the output vertices and applying \texttt{transform}.
Our implementation restricts \texttt{transform} to a matrix multiplication, which we implement using a $16\times32$ weight stationary PE array~\cite{chen2017eyeriss}.
Each PE contains a 16-bit multiplier, as well as a local double buffered weight register.
The PE array is broken into two $16\times16$ blocks.
Blocks can be configured to use one of two modes: cooperative, where both blocks operate on the same vertex, or parallel, where blocks operate on different vertices in parallel.
Parallel mode broadcasts weight values to both blocks, allowing for slightly lower energy consumption at the expense of higher latency when there is only a single output vertex.

Unlike many other neural network accelerators, \name{} does not use a systolic array structure.
Instead, \name{} broadcasts inputs across the rows of the array and accumulates results down columns using a reduction tree.
The entire operation is pipelined to allow multiple matrix operations to occur without stalling, even as weights are transferred in and out of the array.
This results in a significant savings in latency for a single matrix-vector operations; instead of requiring $16+32 = 48$ cycles, our implementation requires just six (three to distribute values, one for multiplication, and two for reduction).
This also eliminates the buffers required for input skewing in a systolic design.

\subsection{Update Unit}
\label{sec:activation}

The update unit iterates over each vector in the vertex accumulator and applies \texttt{activate}.
Our activate PE allows two possible operations: element-wise ReLU and a two-level configurable lookup-table (LUT) that can be used to approximate many activation functions. 
Each LUT level is implemented as a separate table with 33 and 9 entries, respectively.
Both cover overlapping ranges of input: the first level from $-2^a$ to $2^a$, and the second level from $-2^b$ to $2^b$, where $a$ and $b$ are user configurable values.
The LUT entries lineally partition the range, e.g. entry 0 of level 1 corresponds to $-2^a$, entry 1 corresponds to $-2^a + 2^{a+1}/32$, etc.
To perform an activation computation, the input is first converted to a 16-bit fixed point representation with 4-bits of integer precision.
Each level is then checked in series to see if the input falls in its range.
If so, the closest two LUT values are linearly interpolated to produce an output.
If the values overflow the range for both levels, the input is either clamped to the closest value in the second level, or a user configured linear function is used.
Additionally, the overflow behavior can be configured for both positive and negative inputs independently, allowing the implementation of non-symmetric activation functions.
This simple approximation covers a large number of activation functions, including sigmoid, which is required for models such as G-GCN.

\section{Optimizations}
\label{sec:optimizations}

\name{} implements two major GNN optimizations: execution partitioning and vertex-tiling.
Execution partitioning describes a method to split a \name{} program to operate on partitions of a nodeflow, reducing the amount of on-chip memory required.
\name{} supports pipelining operations on different partitions, improving performance.
Vertex-tiling improves the locality of weights the vertex-accumulate phase, reducing the memory bandwidth required by the vertex unit.
Collectively, these optimizations reduce inference latency for \name{} by a significant factor.

\subsection{Execution Partitioning}
\label{sec:partitioning}

\begin{figure}[tb]
    \centering
    \subcaptionbox{\label{subfig:partition-nodeflow}}{
      \includegraphics{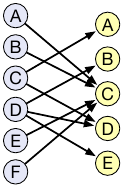}
    }
    \hfill
    \subcaptionbox{\label{subfig:partition}}{
      \includegraphics{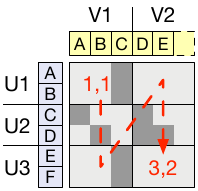}
    }
    \hfill
    \subcaptionbox{\label{subfig:partition-pipeline}}{
      \includegraphics{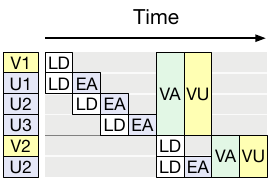}
    }
    \hfill
    \vskip 0.5\baselineskip
    \caption{An example of a nodeflow~\subref{subfig:partition-nodeflow} and corresponding partitions~\subref{subfig:partition} which are processed column-wise. \name{} also pipelines transferring feature data with execution~\subref{subfig:partition-pipeline}.}
    \label{subfig:nodeflow-partition}
\end{figure}

A common GNN optimization is to split the graph into partitions that can be computed on separately~\cite{ma2019neugraph, recoml20greta, yan2020hygcn}.
This reduces the peak amount of on-chip memory required to compute inference since only a portion of the graph must be loaded at once.
\name{} supports a similar optimization we refer to as execution partitioning, shown in Fig.~\ref{subfig:nodeflow-partition}.
First, the user partitions the nodeflow offline by splitting the input and output vertices into fixed chunks of size $N$ and $M$.
Likewise, the edges are partitioned into blocks of size $N \times M$, where block $NF_{i,j}$ stores the edges connecting input vertices in chunk $U_i$ to output vertices in chunk $V_j$.
During inference, \name{} executes edge-accumulate for each partition in a column, skipping blocks that are empty.
Then, \name{} executes the vertex-accumulate and vertex-update phases once, updating values in the corresponding partition of output vertices.
This ensures every incoming edge for each output vertex is processed before vertex-accumulate.

Another advantage of execution partitioning is that operations can be pipelined between partitions.
\name{} implements two kinds of pipelining related to partitioning.
First, \name{} pipelines loading data from off-chip with the edge-accumulate phase.
This allows overlapping execution with bulk loading feature data for an entire partition.
If enough space is available in the nodeflow buffer, \name{} also optionally caches partition feature data loaded during the processing of the first column to avoid reload data while processing later columns.
Second, transferring weights from the global buffer can be pipelined with processing an entire column.
\name{} performs inter-layer pipelining by loading the weights of the next layer while processing the last column, and preloads the tile buffer before processing the first column.

\subsection{Vertex-Tiling}
\label{sec:tiling}

The bandwidth required to load layer weights can be a significant bottleneck.
For example, consider a GCN layer with a feature size of 256.
Since our implementation of \texttt{transform} cannot hold the entire \SI{1}{\mega\byte} weight matrix locally, new weight values must be loaded every cycle.
At an operating frequency of \SI{1}{\giga\hertz}, this requires a maximum of \SI[per-mode=symbol]{2}{\tera\byte\per\second} of tile buffer bandwidth, which we found difficult to implement physically.
While this could be resolved by increasing the number of weights stored within the multiplier array, this increases energy usage and lacks flexibility since a model with a larger feature size would still run into the same limitation.

\begin{figure}[tbh]
    \centering
    \includegraphics{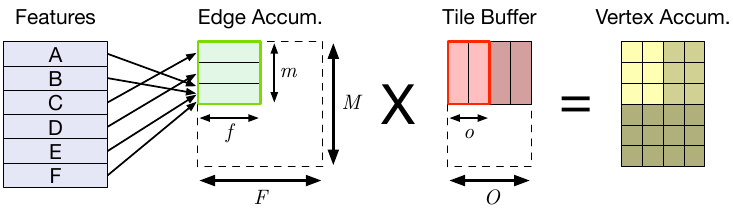}
    \vskip 0.5\baselineskip
    \caption{Vertex-tiling allows materializing a small tile of edge accumulator values ($m\times f$) instead of the full $M\times F$ matrix.
    This reduces the memory bandwidth required since a tile of weight values can be reused across $m$ vertices.}
    \label{fig:tiling}
\end{figure}

\name{}'s approach is to instead use an optimization we call \textit{vertex-tiling}.
The key insight of vertex-tiling is that in almost all cases \texttt{transform} is affine, which allows us to perform an optimization similar to tiling matrix multiplication.
Fig.~\ref{fig:tiling} shows a graphical representation of this strategy.
Here, edge-accumulate produces $f$ elements for $m$ output vertices.
This requires storing $f \times m$ elements in the edge accumulator instead of the full $F \times M$ matrix.
Then, we run vertex-accumulate (in this case matrix multiplication), which loads each $f \times o$ tile from the tile buffer.
We then repeat this process, first for all vertex tiles and then for all weight slices, maximizing the locality of the weights.
This reduces the bandwidth between the tile buffer and the matrix unit by a factor of $1/m$.
Thus, by tuning $f$ and $m$ we can trade-off the required bandwidth with the amount of storage required for tiles and edge-accumulate values.

\section{Experimental Methodology}
\label{sec:experimental-methodology}

\textbf{Datasets}.
Table~\ref{tab:datasets} describes the properties of the datasets chosen for evaluation.
Datasets were selected from previous evaluations of GNNs~\cite{hamilton2017inductive}, the SNAP project~\cite{snapnets}, and the UF sparse matrix collection~\cite{davis2011university}.
Included datasets were designed to be similar to the workloads used by GNNs, as well as provide a range of connectivity.
We prepossessed each dataset using the same procedure outlined by the authors of GraphSAGE~\cite{hamilton2017inductive}.
The column ``2-Hop'' denotes the median number of unique vertices within the 2-hop neighborhood of a vertex picked uniformly at random from the graph, taking into account the sampling procedure.

\begin{table}[thb]
\centering
\caption{Datasets used for evaluation.}
\begin{tabular}{lrrr}
    \toprule
    Dataset & Nodes & Edges & 2-Hop \\
    \midrule
    Youtube (YT) & 1,134,890 & 2,987,624 & 25 \\
    Livejournal (LJ) & 3,997,962 & 34,681,189 & 65 \\
    Pokec (PO) & 1,632,803 & 30,622,564 & 167 \\
    Reddit (RD) & 232,383 & 47,396,905 & 239 \\
    \bottomrule
\end{tabular}
\label{tab:datasets}
\end{table}

\textbf{Models}.
We implemented four GNNs which cover a broad range of different model types: GCN~\cite{kipf2017semi}, the max variant of GraphSage~\cite{hamilton2017inductive}, GIN~\cite{xu2018powerful}, and G-GCN~\cite{bresson2017residual, marcheggiani2017encoding, allamanis2018learning}.
For our neighborhood function, we use the same sampling proceedure as described by the authors of GraphSage.
Specifically, we deterministically map a given vertex to a fixed-sized, uniform sample of its neighbors.
For all models, we use two layers with sample sizes $25$ and $10$ for the first and second layer, respectively.
Samples between layers are independent.
Additionally, we use a feature size of 602 (the feature size of the Reddit dataset), a hidden dimension of 512, and an output dimension of 256 for all layers.

\textbf{Baseline}.
Our CPU baseline was a dual socket server containing two, 14-core \SI{2.60}{\giga\hertz} Intel Xeon E5-2690 v4 CPUs, each with four channels of DDR4-2400 memory.
We restricted our experiments to a single socket to adhere to Tensorflow performance guidelines~\cite{intel2019tfperf} and to avoid latency variation resulting from NUMA.
In this configuration, we measured a sustained \SI[per-mode=symbol]{1.084}{\tera\FLOP\per\second} in a matrix multiply benchmark (93\% of \SI[per-mode=symbol]{1.164}{\tera\FLOP\per\second} theoretical peak) and \SI[per-mode=symbol]{64.5}{\gibi\byte\per\second} of off-chip memory bandwidth (84\% of \SI[per-mode=symbol]{76.8}{\gibi\byte\per\second} theoretical peak).

We implemented both the baseline and our optimized inference algorithm in Tensorflow v2.0~\cite{tensorflow2015-whitepaper} with eager mode disabled and compiled with the Intel Math Kernel Library~\cite{intel2019mkl}.
To discount the overhead of the Tensorflow library for each model, we measured the time to evaluate an equivalent model with all tensor dimensions set to zero and subtract it from the latency measurement.
We also perform a warm-up inference before all measurements to allow Tensorflow to compile and optimize the network.

\begin{table}[tbh]
\centering
\caption{Architectural characteristics of baseline and GRIP.}
\begin{tabular}{lll}
    \toprule
    & CPU & GRIP  \\
    \midrule
    Compute 
        & \makecell[l]{
          \SI[per-mode=symbol]{1.164}{\tera\OP\per\second} \\ @ \SI{2.6}{\giga\hertz}
        }
        & \makecell[l]{
          \SI[per-mode=symbol]{1.088}{\tera\OP\per\second} \\ @ \SI{1.0}{\giga\hertz}
        }
    \\
    \midrule
    \makecell[l]{On-chip\\memory}
        & \makecell[l]{
            L1D: $14\times32$ \si{\kibi\byte} \\ 
            L2: $14\times256$ \si{\kibi\byte} \\ 
            LLC: \SI{35}{\mebi\byte}
        }
        & \makecell[l]{
            Nodeflow: $4\times20$ \si{\kibi\byte} \\
            Tile: $2\times64$ \si{\kibi\byte} \\
            Weight: \SI{2}{\mebi\byte} \\
        }
    \\
    \midrule
    \makecell[l]{Off-chip\\memory}
        & \makecell[l]{$4\times$ DDR4-2400 \\ \SI[per-mode=symbol]{76.8}{\gibi\byte\per\second}}
        & \makecell[l]{$4\times$ DDR4-2400 \\ \SI[per-mode=symbol]{76.8}{\gibi\byte\per\second}}
    \\
    \midrule
    Total Area & \SI{306.18}{\milli\meter\squared} & \SI{11.27}{\milli\meter\squared} \\
    \midrule
    Power & \SI{135}{\watt} & \SI{4.9}{\watt} \\
    \bottomrule
\end{tabular}
\label{tab:arch-overview}
\end{table}

\textbf{ASIC Synthesis}:
We implemented \name{} in SystemVerilog, choosing the architectural parameters to have similar compute and memory bandwidth as our CPU baseline (Table~\ref{tab:arch-overview}).
The implementation uses 16-bit fixed point, which maintains suitable inference accuracy in the models we evaluate.
We then performed synthesis and place and route in a \SI{28}{\nano\meter} CMOS process, targeting a \SI{1}{\giga\hertz} operating frequency and worst case PVT corner.
The critical path of \name{} was determined to be \SI{0.93}{\nano\second}, inside the weight SRAMs.
Power estimates of each unit was performed by generating activity factors from a cycle accurate simulation of our implementation and applying them to our synthesized design.
We used Cacti v6.5~\cite{shivakumar2001cacti} to estimate the area and power of the SRAM memories.
We also integrated Ramulator~\cite{kim2016ramulator} into our simulator to estimate DRAM timings and produce a command trace.
These traces were fed to DRAMPower~\cite{chandrasekar2012drampower} to estimate DRAM power.

\section{Evaluation}
\label{sec:evaluation}

\name{} aims to accelerate GNN inference for a wide range of models, specifically targeting low latency.
We evaluate this by measuring overall inference latency for four different models and compare to a CPU and GPU baseline (Sec.~\ref{sec:eval-performance}).
To better understand \name{}'s performance, we then breakdown the contribution of each architectual feature (Sec.~\ref{sec:eval-breakdown}) and how the overall speedup changes as we modify both architectural (Sec.~\ref{sec:eval-arch-parameters}) and model parameters (Sec.~\ref{sec:eval-model-parameters}).
We also measure the impact of each GNN optimization we implemented (Sec.~\ref{sec:eval-optimization}).
Finally, we compare \name{} to alternative approaches (Sec.~\ref{sec:eval-prior-work}) and present a breakdown of energy consumption during inference (Sec.~\ref{sec:eval-energy}).

\subsection{Overall Performance}
\label{sec:eval-performance}

\begin{table}[t]
\centering
\caption{99\%-ile inference latency for \name{}, CPU, and GPU}
\begin{adjustbox}{width=1\linewidth}
\begin{tabular}{@{}llrrrrr@{}}
\toprule
& & & \multicolumn{2}{c}{CPU} & \multicolumn{2}{c}{GPU} \\
\cmidrule(lr){4-5}
\cmidrule(lr){6-7}
 Model   & Dataset     & GRIP & \si{\micro\second} & $\times$ & \si{\micro\second} & $\times$ \\
\midrule
 \multirow{4}{*}{GCN}
 & youtube     &   15.4 &  309.2 & (20.1) & 1082.4 & (70.5) \\
 & livejournal &   15.8 &  466.8 & (29.5) & 1313.6 & (83.1) \\
 & pokec       &   16.0 &  477.1 & (29.8) & 1085.6 & (67.7) \\
 & reddit      &   16.3 &  407.1 & (25.0) &  813.2 & (50.0) \\
\midrule
\multirow{4}{*}{G-GCN}
 & youtube     &  134.1 & 2315.9 & (17.3) & 1332.5 & ( 9.9) \\
 & livejournal &  146.3 & 2493.2 & (17.0) & 1837.6 & (12.6) \\
 & pokec       &  146.7 & 2637.9 & (18.0) & 1409.2 & ( 9.6) \\
 & reddit      &  147.0 & 2864.2 & (19.5) & 1133.9 & ( 7.7) \\
\midrule
\multirow{4}{*}{GS}
 & youtube     &  113.7 & 1545.1 & (13.6) & 1309.0 & (11.5) \\
 & livejournal &  124.4 & 1947.4 & (15.7) & 2193.8 & (17.6) \\
 & pokec       &  124.9 & 2075.7 & (16.6) & 1759.1 & (14.1) \\
 & reddit      &  125.3 & 2099.0 & (16.8) & 1252.8 & (10.0) \\
\midrule
\multirow{4}{*}{GIN}
 & youtube     &   30.5 &  344.7 & (11.3) & 1387.6 & (45.5) \\
 & livejournal &   30.9 &  416.1 & (13.5) & 1221.5 & (39.5) \\
 & pokec       &   31.1 &  340.7 & (10.9) &  855.5 & (27.5) \\
 & reddit      &   31.4 &  354.8 & (11.3) & 1009.4 & (32.2) \\
\bottomrule
\end{tabular}
\end{adjustbox}
\label{tab:speedups}
\end{table}

To evaluate \name{}'s overall performance, we measured the total end-to-end execution time (latency) to compute inference with a variety of models and datasets. 
Table~\ref{tab:speedups} shows \name{}'s inference latency and speedup versus our CPU and GPU implementation.
We use 99th percentile latency for consistency with prior evaluations of inference performance~\cite{reddi2019mlperf}.

\textbf{Performance vs. CPU}.
Compared to our CPU implementation, \name{} achieves a latency improvement of between $29.8\times$ (GCN, Pokec) and $10.9\times$ (GIN, Pokec) with a geometric mean of $17.0\times$ across all datasets and models.
\name{} tends to give a smaller speedup on models that perform a larger portion of their computation during the \textit{Update} step of the message-passing layer.
For example, GIN's \textit{Update} uses a two-layer MLP that requires roughly double the computation of GCN's single matrix multiplication.
However, the additional computation results in similar overall CPU inference latency since our implementation is largely bottlenecked by non-computational factors (Sec.~\ref{sec:performance-challenges}).
This results in \name{} achieving a smaller performance improvement of 10.9-13.5$\times$ compared to an improvement of more than $13.6\times$ for all other models.

\textbf{Performance vs. GPU}.
Practical deployments of online GNN inference most often use CPUs due to the large memory requirements for graph features and low utilization at small batch sizes.
However, for completeness we also benchmark \name{} against an Nvidia P100 GPU implementation for each model.
\name{}'s speedup on GPU ranges from $83.1\times$ (Livejournal, GCN) to $7.7\times$ (Reddit, G-GCN) with a geometric mean of $23.4\times$.
For models with relatively low overall latency (GCN, GIN) we see a significantly higher speedup than with our CPU implementation.
This is largely due to the overhead of transferring embeddings from host to GPU memory (roughly 200-\SI{500}{\micro\second}, depending on the neighborhood size) which comprises a large portion of the overall execution time for models like GCN (25\%-50\% of total latency).
\name{} does not incur this penalty since features and weights are already stored in device DRAM and do not have to be transferred from the host.
On models with a higher total execution time (e.g. G-GCN), \name{} still achieves a significant speedup due to low GPU utilization.
With a batch size of 1, there is not sufficient computation during each layer to fully utilize the computational resources of the GPU and overhead of launching each kernel tends to dominate.

\subsection{Breakdown of Performance}
\label{sec:eval-breakdown}

In this subsection, we breakdown the performance impact of each architectural feature of \name{}.
Specifically, we modify our cycle-accurate simulator to match the bottlenecks exhibited by our CPU implementation and then progressively remove each modification to measure the impact of different units.
As a performance benchmark, we use the geometric mean speedup of GCN for the largest neighborhood in each dataset.

\textbf{Baseline Configuration}.
Our baseline configuration emulates each core being assigned independent vertices and performing all GReTA phases, with weights and partition data being first loaded into L3 cache and intermediate values accumulated directly in L2.
This results in the following simulator modifications.
First, we modify our vertex unit to use 14, $8\times2$ matrix multiply units, with each unit assigned independent vertices within a partition.
This emulates the effect of each CPU core using two 8-element SIMD units.
Second, we increase the number of fetch and gather units to 14 and the crossbar width to 32 bytes, matching the number of cores and L2 cache bandwidth, respectively.
We also disable pipelining between the edge and vertex units to emulate a single core performing both functions.
Third, we merge the weight and nodeflow buffers into a single SRAM and limit the maximum read bandwidth to 16 bytes per cycle per fetch unit, matching the bandwidth of the L3 cache.
Finally, we disable pipelining between the vertex and update unit to model both operations being performed by the same core.
This configuration overestimates the performance of the CPU in practice since it models ideal performance and no additional computation required for auxiliary operations, such as indexing calculations.
In particular, with a \SI{2.6}{\giga\hertz} clock and an element width of 4-bytes, our model is $2.07\times$ faster than the measured CPU latency.

\begin{figure}
  \centering
  \subcaptionbox{Speedup breakdown for each component of \name{} versus baseline.\label{subfig:breakdown-stacked}}{
    \includegraphics{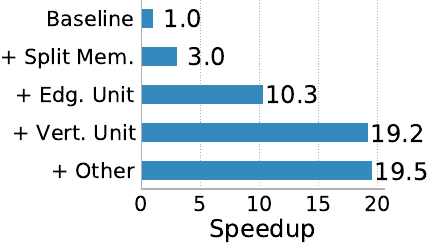}
  }
  \hfill
  \subcaptionbox{Estimated speedups of prior work versus baseline and \name{}.\label{subfig:prior-work}}{
    \includegraphics{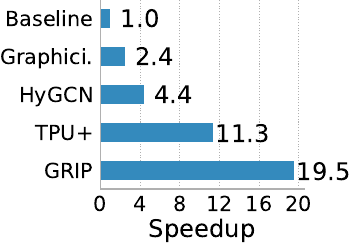}
  }
  \vskip \baselineskip
  \caption{Breakdown of performance improvements.}
  \label{fig:breakdown}
\end{figure}

\textbf{Breakdown}.
In Fig.~\ref{subfig:breakdown-stacked}, we show the impact of different units in \name{} by progressively removing each modification from our baseline in reverse order. 
First, we split the weight and nodeflow memories into separate SRAMs.
This results in a 2.8$\times$ speedup due to removing contention between fetching features and weights from the same SRAM ($2.0\times$), as well doubling the bandwidth available to load weight values into the vertex unit ($1.4\times$).
Second, we add the edge unit, resulting in an additional improvement of $3.4\times$.
While this is partially due to increased crossbar bandwidth after adjusting the number of fetch and gather units ($1.14\times$), the majority of the speedup is due to allowing loading data, edge-accumulate, and vertex-accumulate phases to overlap by using a dedicated unit for each phase ($2.97\times$).
Third, we enable the vertex unit and revert to using a single $16\times32$ matrix multiply unit, resulting in an additional $1.87\times$ speedup.
This is due to increased overall TOP/s ($1.63\times$) and using a single unit rather than multiple units, which allows units to not be wasted when the overall number of output vertices is small ($1.15\times$).
Finally, separating and pipelining the update unit produces a small speedup of $1.02\times$.

\subsection{Architectural Parameters}
\label{sec:eval-arch-parameters}

Here, we discuss the impact of several high level architectural parameters on inference performance.

\begin{figure}
  \centering
  {
    \includegraphics{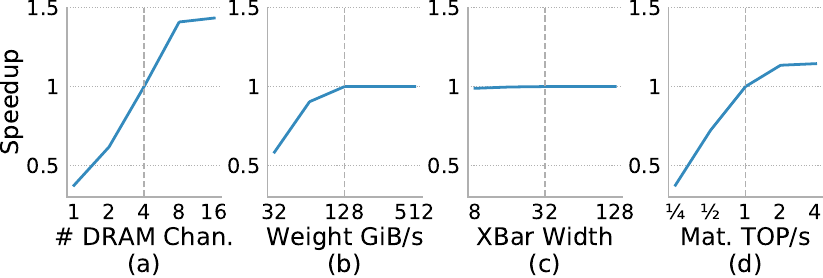}
    \phantomsubcaption\label{subfig:scalability-dram}
    \phantomsubcaption\label{subfig:scalability-weight}
    \phantomsubcaption\label{subfig:scalability-xbar}
    \phantomsubcaption\label{subfig:scalability-mat}
  }
  \caption{Impact of scaling architectural parameters. Dashed vertical line indicates our implementation's parameters. In Fig.{~\ref{subfig:scalability-dram}} the number of edge unit lanes is kept equal to the number of channels.}
  \label{fig:scalability}
\end{figure}

\textbf{Number of DRAM Channels}.
The number of DRAM channels determines the overall memory bandwidth available to transfer data on- and off-chip.
In Fig.~\ref{subfig:scalability-dram}, we observe that \name{}'s performance is strongly related to the number of channels until around 8 channels ($\sim$\SI[per-mode=symbol]{150}{\gibi\byte\per\second}).
This indicates that \name{}'s performance is primarily limited by off-chip memory bandwidth.

\textbf{Weight Bandwidth}.
The weight bandwidth determines how many values can be read from the global weight buffer each cycle.
If this is set too low, loading weight values can become a bottleneck during vertex-accumulate.
We observe this effect in Fig.~\ref{subfig:scalability-weight} below \SI[per-mode=symbol]{128}{\gibi\byte\per\second}, which corresponds to loading 64 weight values each cycle.

\textbf{Crossbar Port Width}.
The crossbar port width determines the number of elements accumulated by each gather unit in a single cycle.
In our experiments, the average number of edges per vertex is fairly small (sampled to be less than 25).
Since edge-accumulate typically takes much less time than vertex-accumulate or loading data from DRAM, increasing the width has a limited impact on performance (Fig.~\ref{subfig:scalability-xbar}).
However, it is preferable to over-allocate the crossbar width in order to ensure high performance even on dense nodeflows.

\textbf{Matrix Multiply TOP/s}.
The total number of TOP/s \name{} can achieve is determined primarily by the size of the matrix multiply unit.
In Fig.~\ref{subfig:scalability-mat} we see that performance is strongly related to the size of this unit, until reaching around \SI[per-mode=symbol]{2}{\tera\OP\per\second} at which point \name{} is limited by memory bandwidth.
Thus, our implementation of \name{} would see a relatively small benefit from a substantially larger matrix unit ($1.14\times$ for a $4\times$ larger unit).

\subsection{Model Parameters}
\label{sec:eval-model-parameters}

\begin{figure}
  \centering
  \subcaptionbox{Impact of feature dimensions.\label{subfig:feature-dims}}{
    \includegraphics{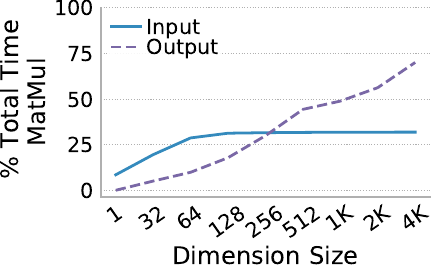}
  }
  \hfill
  \subcaptionbox{Impact of sampling.\label{subfig:sampling}}{
    \includegraphics{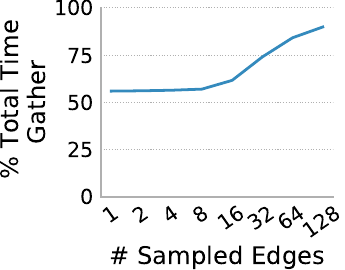}
  }
  \vskip \baselineskip
  \caption{The impact of scaling different GCN parameters on the balance of time spent in each operation.
  Scaling the output feature size increases the amount of time spent performing matrix multiplication, while increasing the number of edges decreases it.}
\end{figure}

A key aspect \name{}'s design is balancing the performance between GReTA's edge and vertex-centric phases.
Here, we evaluate how this balance changes as the parameters of the GNN model are altered.

\textbf{Feature Dimensions}.
In Fig.~\ref{subfig:feature-dims}, we evaluate how varying the number of the input and output features impacts the percent of time spent in matrix multiplication.
The proportion is initially low ($\sim$8\%) for small features and increases linearly until 32 features.
This is due to the fact that when the feature size is smaller than the native width of the DRAM interface, DRAM bandwidth is poorly utilized due to many random accesses.
In our implementation, we use two dual-channel DRAM controllers, which each have an interface of 64 2-byte elements.
Above this point, the proportion of time spent performing vertex-accumulation stays flat, reflecting the fact that each additional feature results in a constant amount of additional computation during inference.
However, this analysis does not hold for the output features, which can be increased without needing to increase the number of values loaded from DRAM.
We see that increasing the output feature size always increases the percent of time performing vertex-accumulate.
Thus, models with large output feature sizes are likely to be limited by compute rather than memory.

\textbf{Sampled Edges}.
Another important model parameter is the number of sampled edges per output vertex.
In Fig.~\ref{subfig:sampling}, we evaluate how the number of edges impacts the percent of total time spent performing edge-accumulate.
For less than 8 edges per vertex, \name{}'s performance is mostly limited by computation and overhead related to accessing data from DRAM.
Above this threshold, the memory and crossbar bandwidth becomes a bottleneck, and \name{} spends an increasing portion of execution loading data.

\begin{figure}[tbh]
    \centering
    \subcaptionbox{\label{subfig:latency-by-size}}{
        \includegraphics{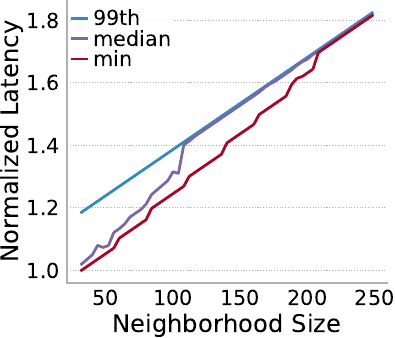}
    }
    \hfill
    \subcaptionbox{\label{subfig:speedup-by-size}}{
        \includegraphics{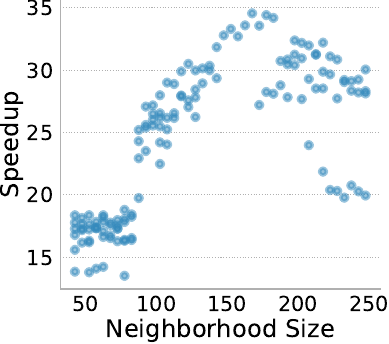}
    }
    \vskip \baselineskip
    \caption{Impact of different neighborhood sizes on latency for the GCN model. \name{}'s latency linearly increases with neighborhood size due to more computation being required for inference. The speedup is roughly constant until a neighborhood size of about 95, at which point intermediate values no longer fit into the cache of a single CPU core.}
    \label{fig:neighborhood_impact}
\end{figure}

\textbf{Neighborhood Size}.
The neighborhood size heavily impacts \name{}'s overall latency and is influenced by local graph structure.
To demonstrate the impact of the neighborhood size on performance, we plot \name{}'s minimum, median, and 99th percentile inference latency for GCN across different neighborhoods of the LiveJournal dataset in Fig.~\ref{subfig:latency-by-size}.
The result is a strong linear relationship between the neighborhood size and latency across the entire distribution.
Each vertex added to the neighborhood results in a roughly constant increase in the amount of work during inference.
Additionally, we observe that as the neighborhood size increases, the median latency moves closer to the 99th percentile.
This is the result of larger neighborhoods being more likely to be densely connected, leading to a larger number of reductions that must be computed.

In Fig.~\ref{fig:neighborhood_impact}, we evaluate the latency speedup compared to the CPU baseline across different neighborhood sizes.
Below a neighborhood size of 95, we see a roughly constant speedup of between {12$\times$} and {18$\times$}.
For these neighborhood sizes, all intermediate values fit into the L1 and L2 cache of a single CPU core.
After this point, some feature values must be stored in the L3 cache and inference performance becomes limited by the cache bandwidth (Sec.~\ref{sec:performance-challenges}).

\subsection{Optimizations}
\label{sec:eval-optimization}

\begin{figure}[thb]
  \centering
  \subcaptionbox{Impact of pipelining\label{subfig:pipeline-breakdown}}{
    \includegraphics{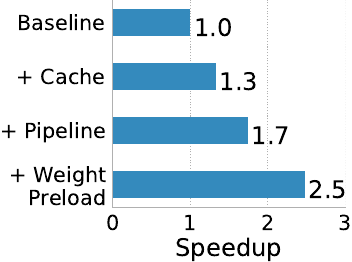}
  }
  \hfill
  \subcaptionbox{Speedup of vertex-tiling\label{subfig:tiling-heatmap}}{
    \includegraphics{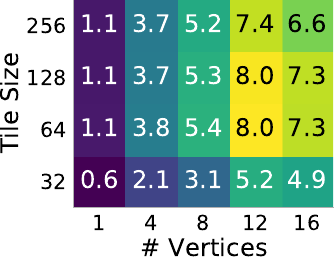}
    \hspace{0.3in}
  }
  \vskip \baselineskip
  \caption{Impact of partitioning and tiling optimizations.}
\end{figure}

In this subsection, we evaluate the impact of each optimization used by \name{}.

\textbf{Partitioning and pipelining}.
In Fig.~\ref{subfig:pipeline-breakdown} we show the cumulative speedups of each optimization enabled by partitioning.
We compare to an unoptimized baseline, where feature values are loaded from off-chip on demand and no pipelining exists between stages.
First, by caching feature data on-chip, \name{} achieves a $1.3\times$ speedup.
This is due to the decreased memory traffic required to reload data between partition columns and by the improved throughput from bulk loading data for an entire partition.
Second, pipelining operations between different partitions results in an additional $1.3\times$ speedup due to overlapping execution with memory transfers. 
Finally, we can also pipeline the transfer of weights from the global weight buffer into the vertex-update unit.
This increases the overall speed-up to a total of $2.5\times$.

\textbf{Vertex-Tiling}.
In Fig.~\ref{subfig:tiling-heatmap}, we show the speedup compared to no tiling as we alter the two tiling parameters $M$ (the number of vertices in a tile) and $F$ (the number of elements per vertex).
We see that performance generally reaches a maximum around $F=64$ elements.
Above $F=64$, increasing $F$ causes the vertex unit to stall more often while waiting for a tile to be produced by the edge unit.
The performance degradation is not linear because the time taken to accumulate a tile depends on the connectedness of the nodeflow.
We also see degraded performance below $F=64$.
This is because $F$ features are loaded from memory for each vertex.
As $F$ decreases, more random DRAM accesses are required to load features, degrading DRAM throughput.
Increasing $M$ also increases performance until around 12 vertices.
The maximum number of output vertices in our model is 11.
Increasing $M$ beyond this only adds additional latency associated with processing empty dummy vertices.

\subsection{Comparisons to Prior Work}
\label{sec:eval-prior-work}

Several other approaches have been proposed to accelerate neural networks and graph algorithms.
Here, we analyze the bottlenecks present in each approach and compare performance with \name{}.

\textbf{HyGCN}.
HyGCN~\cite{yan2020hygcn} is an accelerator designed for graph convolutional networks, a subset of GNNs that do not have computation associated with edges.
HyGCN and \name{} take a similar approach of using separate units for edge- and vertex-centric operations.
However, \name{} addresses two major bottlenecks present in the HyGCN design.

First, HyGCN uses 32 8-lane SIMD units to perform edge-oriented operations, but can only issue a single edge at a time.
This means the throughput of edge operations will be limited when the number of features is smaller than the total number of SIMD lanes.
In contrast, \name{} allows for multiple edges to be issued in parallel.

Second, HyGCN requires an entire feature vector to be computed and stored before performing vertex-oriented operations.
In order to process multiple vertices in parallel, this requires a large buffer to store accumulated values ({\SI{16}{\mega\byte}} in the HyGCN implementation).
The size of this buffer is also reported by the HyGCN authors to have a significant impact on their overall performance ($1.3$ -- $4 \times$ worse performance for a $16 \times$ smaller buffer.)
In contrast, \name{} uses vertex-tiling to only store a small number of elements from multiple feature vectors.
This allows \name{} to use a roughly $10,000\times$ smaller buffer ({\SI{1.5}{\kibi\byte}) while achieving comparable performance.}

We demonstrate these limitations by modifying our simulator to emulate the HyGCN approach.
Specifically, we set the number of gather and fetch units to 1 and the crossbar width to 256 to match the number of SIMD lanes.
We disable all tiling and force feature vectors to be fully accumulated before vertex-accumulate. 
We then set all other parameters to be the same as \name{}, including the same partitioning used in our evaluation of \name{}.

This configuration results in a speedup of $4.4\times$ the baseline, shown in Fig.~\ref{subfig:prior-work}.
However, it performs $4.5\times$ slower than \name{} due to limits in the available on-chip memory bandwidth for weights.
Incorporating vertex tiling would allow for a much smaller edge accumulate buffer and reduce the required bandwidth by increasing the reuse of the weights.

\textbf{Modified TPU}.
The TPU~\cite{jouppi2017datacenter} is a DNN accelerator designed around a large 2-D systolic array.
Unfortunately, GNNs are difficult to implement efficiently for the TPU due to a lack of support for edge-oriented operations~\cite{balog2019fast}.
Instead, we compare \name{} to a modified version of the TPU architecture that addresses this limitation by incorporating features from \name{}.
We refer to this modified design as the TPU+.

Specifically, the TPU+ has an additional unit similar to \name{}'s edge-unit between the TPU's unified buffer and the systolic data setup.
This allows the TPU+ to natively support the GReTA programming model by mapping \textit{edge-accumulate} onto the new edge-unit,  \textit{vertex-accumulate} onto TPU's systolic array, and \textit{vertex-update} onto the activation pipeline.
This design also supports both the execution partitioning and vertex-tiling optimizations described in Sec.~\ref{sec:optimizations}.

We estimate the performance of the TPU+ by modifying our cycle-accurate model to use a single fetch and gather unit.
We also replace the vertex-unit with an identically sized $16 \times 32$ systolic array.
As in the original TPU design, weights are stored off-chip and the dedicated weight bandwidth is limited to \SI[per-mode=symbol]{30}{\gibi\byte\per\second}.
All other parameters remain unchanged compared to our evaluation of \name{}, including the use of $4\times$ DDR4-2400 for off-chip memory and the same partitioning and vertex-tiling optimizations.

This configuration achieves a $11.3\times$ speedup (Fig.~\ref{subfig:prior-work}) compared to our baseline in Sec.~\ref{sec:eval-breakdown}.
The main bottleneck in this approach is the limited bandwidth dedicated to weights.
Moving weights on-chip as in \name{} results in a $1.72\times$ speedup.
Higher performance memory for weights (e.g. HBM as used by later versions of the TPU) could also address this bottleneck.
However, we leave a fuller exploration for future work.

\textbf{Graphicionado}.
Graphicionado~\cite{ham2016graphicionado} is an accelerator architecture designed for graph analytics.
Like \name{}, Graphicionado allows several units to be specialized for a particular algorithms, such as GCN inference.
However, it is designed for algorithms that use a small amount of state per-vertex.
As a result, it suffers from two bottlenecks.
First, like HyGCN, it cannot perform vertex-tiling since it requires full feature vectors to be accumulated.
This results in a bottleneck similar to HyGCN since weight data cannot be easily reused between different vertices.
Second, each lane has independent vertex units instead of using a single shared unit, increasing the required weight bandwidth by an amount proportional to the number of lanes.

We estimate the impact of these bottlenecks by modifying our simulator by disabling tiling and splitting the vertex unit into two units lanes that share a single tile buffer port.
We also use the same partitioning scheme used for \name{}.
This configuration results in a small speedup of $2.4\times$ over the baseline, shown in Fig.~\ref{subfig:prior-work}.
However, this is $8.1\times$ slower than \name{} due to significant bottlenecks in weight bandwidth.

\subsection{Energy}
\label{sec:eval-energy}

Table~\ref{tab:energy} shows the power consumption for each of \name{}'s core top level modules during GCN inference.
The single most energy intensive during inference is loading embeddings from DRAM, consuming more than the rest of the accelerator combined (53.7\%).
This is due to the fact that both the number of vertices and the feature size is the largest at the input of GCN, leading to more data being initially loaded in the first layer.
Additionally, \name{} optimizes for latency with four high performance DRAM channels, requiring a large amount of energy per transfer.
The rest of the energy is mostly used by loading weights from the global weight and nodeflow buffers.
Both are fairly large, leading to a high energy cost per read and write.
In total, \name{} uses just \SI{4.9}{\watt}, a significant improvement over the \SI{135}{\watt} TDP of the baseline CPU.

\begin{table}[thb]
\centering
\caption{Breakdown of power for GCN inference.}
\begin{adjustbox}{width=0.7\linewidth}

\begin{tabular}{@{}cccrr@{}}
\toprule
& Module & \phantom{a} & \si{\milli\watt} & (\%) \\
\midrule
\multirowcell{3}{Execution\\Units}
& Edge && 4.1 & 0.1 \\
& Vertex && 656.6 & 12.6 \\
& Update &&  0.4 & $<0.1$ \\
\midrule
\multirow{2}{*}{SRAM}
& Weight && 1476.7 & 28.3 \\
& Nodeflow && 269.5 & 5.1 \\
\midrule
DRAM & - && 2794.7 & 53.7 \\
\midrule
 & Total && 4932.4 & 100 \\
\bottomrule 
\end{tabular}
\end{adjustbox}
\label{tab:energy}
\end{table}



\section{Conclusion}
\label{sec:conclusion}

GNNs represent a promising new method in machine learning to learn directly from graph-structured data.
However, the computational costs of GNNs represent a significant barrier for deployment in many applications, especially in the scenario of online inference.

This paper presents \name{}, an accelerator architecture designed for low latency GNN inference.
\name{} splits GNN operations into a series of edge- and vertex-centric phases.
Each phase is implemented independently in hardware, allowing for specialization of both the memory subsystem and execution units to improve performance.
Additionally, \name{} has hardware support for several optimizations that further reduce latency, including pipelining operations between nodeflow partitions and vertex-tiling.
We then implement \name{} as \SI{28}{\nano\meter} ASIC capable of executing a range of different GNNs.
On a variety of real graphs, our implementation improves 99th percentile latency by a geometric mean of {$17\times$} and {$23\times$} compared to a CPU and GPU baseline, respectively, while drawing only \SI{5}{\watt}.


\bibliographystyle{IEEEtranS}
\bibliography{refs}

\end{document}